\documentclass[12pt]{article}
\usepackage{epsfig}
\begin{document}
\title{\bf Spectral properties of distance matrices}
\author{E. Bogomolny, O. Bohigas, and C. Schmit\\
 Laboratoire de Physique Th\'eorique et Mod\`eles Statistiques\\
Universit\'e de Paris XI, B\^at. 100\\
91405 Orsay Cedex, France}

\maketitle

\begin{abstract}
Distance matrices are matrices whose elements are the relative distances 
between points located on a certain manifold. In all cases considered here 
all their eigenvalues except one are non-positive. When the points are 
uncorrelated and randomly distributed we investigate the average
density of their eigenvalues  and the structure of their eigenfunctions.
The spectrum exhibits delocalized and strongly localized  states which
possess different power-law average behaviour. The exponents depend only on
the dimensionality of the manifold.
\end{abstract}

\section{Introduction}

\indent

In a recent work about general properties of complete metric spaces
\cite{Vershik} A.M. Vershik introduced a specific type of random matrices,
which he called distance matrices, and asked questions about their
statistical properties.

Distance matrices are defined for any metric space $X$ with some
probability measure on it in the following way. Consider
$N$ points randomly distributed on $X$ according to the  measure.
The matrix element $M_{ij}$ of the $N\times N$ (real symmetric) distance matrix
$M$  equals the distance on $X$ between points $i$ and $j$. In all
cases considered here it is tacitly assumed that there always exists a distance
$||\ldots ||$ on $X$ between 
two points $i$ and $j$ which depends only on their relative position and we
use the notation
\begin{equation}
M_{ij}=||\vec{x}_i-\vec{x}_j||,
\label{mij}
\end{equation}
where $\vec{x}_i$ is the $d$-dimensional vector locating the point $i$ on $X$,
and $d$ is the dimension of the base manifold.

For any realization of the random points,   (real) eigenvalues
$\Lambda_n$ and eigenvectors $u^{(n)}$ of distance matrices are
well defined
\begin{equation}
\sum_{j=1}^{N}M_{ij}u_j^{(n)}=
  \sum_{j=1}^{N}||\vec{x}_i-\vec{x}_j||u_j^{(n)}=\Lambda_n u_{i}^{(n)}.
\label{eigenvalue}
\end{equation}
We are interested in their statistical properties.

The first quantity to be considered is the average eigenvalue density defined as
\begin{equation}
\rho(\Lambda)=<\frac{1}{N}\sum_{j=1}^N\delta (\Lambda-\Lambda_j)>,
\end{equation}
or equivalently the average integrated eigenvalue density, i.e. 
the average staircase function
\begin{equation}
{\bf N}(\Lambda)=<\frac{1}{N}\sum_{j=1}^N\Theta (\Lambda-\Lambda_j)>.
\end{equation}
Here $<\ldots >$ denotes an average taken over realizations.
$N{\bf  N}(\Lambda)$ (counting function) counts the number of eigenvalues
up to the value $\Lambda$.

As the elements of distance matrices are non-negative there is one large
positive eigenvalue whose existence follows from the Perron-Frobenius
theorem (see e.g. \cite{Gantmacher} V. 2, p. 49). When the metric space $X$
is Euclidean or spherical all other eigenvalues
have the remarkable property of being non-positive (the counting function
associated to a distance matrix satisfies ${\bf  N}(0^{+})=(N-1)/N$).  
The proof of this for Euclidean spaces is given in  \cite{Schoenberg} (see
\cite{BBS} for a general discussion of the subject).  
The purpose of the present  note is the investigation
of  the asymptotics, in the limit of a large number of points $N$, of the
average eigenvalue  density at large and small negative eigenvalues and the 
properties of the corresponding eigenfunctions for distance matrices built
from  a uniform distribution of uncorrelated points on a base manifold. 

The plan of the paper is the following.  In Section \ref{section2}
one-dimensional spaces are  considered
in detail. In Section \ref{line} we demonstrate that  the
case of the  interval is equivalent to the one-dimensional Anderson model with
diagonal disorder. Though all states are localized, the localization length
increases for large negative eigenvalues as  discussed in Section
\ref{localization}. When the localization length is much larger than the
system size, the concept of  localization becomes meaningless and  a plane wave
description of such states is adequate. This
happens for large negative eigenvalues and as shown in Section
\ref{crystal} it leads to a power-law behaviour of
the eigenvalue density. For small negative $\Lambda$, states are strongly
localized and in Section \ref{strongly} it is demonstrated that in the
one-dimensional case the eigenvalue density tends to a constant. The properties of the 
participation ratio are also discussed in this Section.  

When instead of the interval  the circle is considered, 
two new phenomena appear. First, as demonstrated in Section
\ref{matcircle}, the delocalized eigenvalues corresponding to large 
negative $\Lambda$ form quasi-doublets whose splittings are much smaller than
the distance among them. Second, as shown in Section \ref{stronglycircle},
the localized eigenfunctions of the distance matrix on the circle  are, in general, 
localized not in one   but in two diametrically opposite regions
(forming a kind of echo). 

In Section \ref{higher} $d$-dimensional spaces are investigated.
First in Section \ref{continuous} we introduce the continuous approximation valid for large negative
eigenvalues and show that it leads to a $|\Lambda |^{-(2d+1)/(d+1)}$ asymptotics
of the average eigenvalue  density.  In Section \ref{multiplets}  it is demonstrated that if the base
manifold has a symmetry group, large negative eigenvalues of its distance
matrix form quasi-multiplets whose dimensions equal the dimensions of
the irreducible representations of the group. For small $|\Lambda |$, the
splitting of these multiplets becomes comparable to the distance among
them and the  quasi-multiplet structure disappears. The general condition
for the applicability of the continuous approximation is discussed in Section
\ref{condition} where it is demonstrated that the quasi-multiplets are
present only for the first $\sqrt{N}$
largest negative eigenvalues. In Section \ref{stronglygeneral}  the
behaviour of the average eigenvalue density for strongly localized states is
investigated and it is shown that it vanishes as $|\Lambda |^{d-1}$. To
investigate eigenfunction properties the participation ratio is considered
in the same Section. The presence of a localization echo is also established
for higher dimensions. 

Numerical calculations when the points are
distributed uniformly  on sphe\-res and cubes of different dimensions are 
consistent with the results found.

\section{One-dimensional spaces}\label{section2}

\subsection{Distance matrices on an interval}\label{line}

Let us consider $N$ uncorrelated points $x_j$  distributed uniformly on an 
interval of length $L$. The distance matrix in this case is 
\begin{equation}
  M_{ij}=|x_i-x_j|,
\label{distline}  
\end{equation}
where $|\ldots|$ stands for the usual modulus.
The eigenvalue equation (\ref{eigenvalue}) reads
\begin{equation}
\sum_{j=1}^{N}|x_i-x_j|u_j=\Lambda u_i, \;\;\mbox{for}\;i=1,\dots,N. 
\label{4}
\end{equation}
The eigenvalues of distance matrices are insensitive to the ordering of the
$N$ points but the understanding of the structure of the eigenvectors depends heavily on it.
We will rearrange the points $x_j$ in increasing
order
\begin{equation}
0\leq x_1 \leq x_2\leq \ldots \leq x_N \leq L.
\label{order}
\end{equation}
\noindent
Subtract Eqs.~(\ref{4}) with indices $i+1$ and $i$ (assuming (\ref{order})), 
then
\begin{equation}
\Lambda (u_{i+1}-u_i)=\sum_{j=1}^N(|x_{i+1}-x_j|-|x_{i}-x_j|)u_j.
\end{equation}
As
\begin{equation}
|x_{i+1}-y|-|x_{i}-y|=(x_{i+1}-x_i)\left \{ \begin{array}{rl} 
-1 &\mbox{when}\; y\geq x_{i+1}\\
1  &\mbox{when}\; y\leq x_{i} \end{array}\right .,
\end{equation}
one gets
\begin{equation}
\Lambda (u_{i+1}-u_i)=(x_{i+1}-x_i)[ -\sum_{j=1}^i u_j+\sum_{j=i+1}^N u_j].
\label{6}
\end{equation}
After simple manipulations one  proves that these equations are equivalent
to 
\begin{equation}
\Lambda (R_{i+1}-2R_i+R_{i-1})=2(x_{i+1}-x_i)R_i,
\label{14}
\end{equation}
where $R_i=L_i-L_N/2$ with $L_i=\sum_{j=1}^i u_j$ for $i=1,\ldots, N$
and $L_0=0$.

This second order difference equation has to be completed with boundary conditions. The
first follows from the definition of $R_i$
\begin{equation}
R_N=-R_0.
\end{equation}
The second one can be obtained from any of Eqs.~(\ref{4})
by expressing $u_i$ through $L_i$. Combining Eqs.~(\ref{4}) with $i=1$ and
$i=N$ one gets
\begin{equation}
\Lambda (-R_1-2R_N+R_{N-1})=2(x_1-x_N)R_N.
\end{equation}
This condition can be casted in  the form of Eqs.~(\ref{14}) by introducing
the point $x_{N+1}=x_1$. Then Eqs.~(\ref{14}) are valid for all 
$i=1,\ldots ,N+1$ 
and the boundary conditions correspond to the anti-symmetric solutions
\begin{equation}
R_N=-R_0,\;\;R_{N+1}=-R_1.
\label{21}
\end{equation}
Eqs.~(\ref{14}) coincide with the one-dimensional Anderson model 
\begin{equation}
R_{i+1}-(E-V_i)R_i+R_{i-1}=0,\;\mbox{for}\;i=1,\ldots,N+1,
\label{22}
\end{equation}
with diagonal disorder 
\begin{equation}
E-V_i=2(1+\frac{l_i}{\Lambda}),
\label{23}
\end{equation}
where $l_i$ $(=x_{i+1}-x_i)$ are random variables equal to the distance between
adjacent points. When $N\rightarrow \infty$ and the points $x_j$ are
uncorrelated $l_i$ are independent random variables with the Poisson 
distribution
\begin{equation}
P(l)=\bar{\rho} \exp (-\bar{\rho} l),
\label{poisson}
\end{equation}
where 
\begin{equation}
\bar{\rho}=\frac{N}{L}
\end{equation}
is the mean density of initial points.

\subsection{Localization length}\label{localization}

It is well known (see e.g. \cite{Pastur}) that all solutions of the
one-dimensional Anderson model (\ref{22}) are exponentially localized i.e.
they have asymptotically the following decay from their maximum value, say
at $n_0$  
\begin{equation}
|R_n|\sim e^{-|n-n_0|/l_{loc}},
\end{equation}
where $l_{loc}$ is the dimensionless localization length.

When $|\Lambda| \rightarrow \infty$ the fluctuating part of the random potential
(\ref{23}) tends to zero and it is convenient to use the perturbation
theory developed in \cite{Derrida}.  The first terms of the expansion of the 
localization length for the one-dimensional Anderson model (\ref{22}) with
a random potential $\epsilon V$ of zero mean ($<V>=0$) are
\begin{equation}
\frac{1}{l_{loc}} \approx \epsilon^{2/3}[\sqrt{x}-\frac{<V^2>}{8x}],
\end{equation}
where $E-2=\epsilon^{4/3}x$.
In our case $\epsilon=1/(\Lambda\bar{\rho})$, $V_i=2(l_i\bar{\rho}-1)$, and
$x=2\epsilon^{-1/3}$. By introducing the dimensionless scaled
eigenvalue
\begin{equation}
\lambda=\bar{\rho}\Lambda=\frac{N}{L}\Lambda ,
\end{equation}  
one has
\begin{equation}
\frac{1}{l_{loc}}  \approx   \sqrt{\frac{2}{\lambda}}-\frac{1}{4\lambda}.
\end{equation}
This expression is valid for positive  $\lambda$. When $\lambda$ is negative
the first term is imaginary and  only the second term remains
\begin{equation}
l_{loc} \rightarrow   -4\lambda,\;\mbox{when}\;\lambda \rightarrow -\infty.
\label{asloc}
\end{equation}

\subsection{Crystal configuration}\label{crystal}

Though for the model (\ref{22})  all states are formally localized, 
only $N$ sites exist in our
problem and, as usual for finite systems,  the
effect of localization can  be   ignored when 
the change of the wave function over the system size is small
\begin{equation}
\frac{N}{l_{loc}} \ll 1.
\end{equation}
For large $N$ Eq.~(\ref{asloc})  indicates that states with $|\lambda |\geq N/4$ are
practically delocalized and all states with smaller $|\lambda |$  are
localized. 

For delocalized states, the fluctuating part of the potential (\ref{23}) is
unimportant. Neglecting it  is equivalent  to investigate the spectrum of
the distance matrix for an equally spaced points configuration
\begin{equation}
x_i=\frac{i}{N+1}L,\;\mbox{for}\;i=1,\ldots,N,
\label{xi}
\end{equation}
which we call the crystal configuration. From Eq.~(\ref{22})  it
follows that for this configuration the $R_i$ take an especially simple
form
\begin{equation}
R_i=aq^i+bq^{-i},
\end{equation}
where $q$ is related to the scaled eigenvalue $\lambda$
\begin{equation}
\lambda=\frac{2q}{(1-q)^2}.  
\end{equation}
The allowed values of $q$ (and consequently of $\lambda$) are determined from
the boundary conditions (\ref{21}). Straightforward calculations give two
equations for $q$
\begin{eqnarray}
q^N+1&=&0,
\label{q1}\\
q^{N+1}+1&=&\frac{N+1}{N-1}(q^N+q).
\label{q2}
\end{eqnarray}
The set of solutions of both  equations (except $q=-1$) corresponds to
eigenvalues of the crystal distance matrix. If $q$ is a solution then $1/q$
is also a solution and both give the same eigenvector and eigenvalue and
the total number of solutions is $N$, as it should be. 

These equations have  only one real solution  which
gives the Perron-Frobenius  (i.e. the largest positive) eigenvalue. All other solutions have the
form $q=\exp (i\phi)$ and correspond to a negative value of $\lambda$
\begin{equation}
\lambda=-\frac{1}{2\sin^2 \phi/2}.
\end{equation}
For large $N$, solutions of Eqs.~(\ref{q1}) and (\ref{q2})  have the form 
\begin{equation}
q=\exp(u/N)
\end{equation}
with $u$ independent on N. The corresponding eigenvalues are
\begin{equation}
\lambda= \frac{1}{2\sinh^2 (u/2N)} \rightarrow \frac{2N^2}{u^2}, 
\;\mbox{when}\;N\rightarrow \infty.
\end{equation}
In this limit Eq.~(\ref{q2}) takes the form
\begin{equation}
\cosh z=z\sinh z
\label{q3}
\end{equation}
where $z=u/2$. Its unique positive real solution is $z\approx 1.19968$ and
the Perron-Frobenius eigenvalue $\lambda \sim 1.6671 N^2$. The imaginary
solutions $u=i\phi$  of Eq.~(\ref{q2}) lead to the following asymptotics
$\phi =2(\pi n -1/(\pi n)) +O(n^{-2})$. Together with the 
solutions of Eq.~(\ref{q1}) $\phi=(2m+1)\pi$ the allowed values of $\phi$
are  approximately
$\phi_n\rightarrow \pi n,\;\mbox{with}\;1\ll n \ll N$ 
and the corresponding scaled eigenvalues are
\begin{equation}
\lambda_n=-\frac{2N^2}{\pi^2 n^2}.
\label{smalln}
\end{equation}
The counting function  ${\bf N}(\lambda)$  of large negative eigenvalues is
\begin{equation}
{\bf N}(\lambda)=\frac{1}{N}\sum_{n=1}^N\Theta(\lambda-\lambda_n)\rightarrow
\frac{C}{(-\lambda)^{1/2}},
\label{delocal}
\end{equation}
with $C=4\sqrt{2}/\pi$. 

The $N\rightarrow \infty$ behaviour 
of the crystal configuration distance matrix can also
be obtained without the knowledge of the exact solution by
noticing that in this
limit the eigenvector components $u_j$ can be replaced by a continuous
function $u(x)$ with $x$ as in Eq.~(\ref{xi}).
In this approximation (which we call the continuous approximation) 
the eigenvalue equation (\ref{4}) takes the form
\begin{equation}
\Lambda u(x)=\frac{N}{L}\int_0^{L}|x-y|u(y)dy.
\label{continue}
\end{equation}
Differentiating this equation twice and taking into account that $|x|''=2\delta (x)$, one gets
\begin{equation}
\Lambda u''(x)=2\frac{N}{L}u(x).
\end{equation}
Its general solution is $u(x)=ae^{v x}+be^{-v x}$, with $\Lambda=2N/(v^2 L)$.
Substituting in Eq.~(\ref{continue}) one obtains
equations for $a$ and $b$ whose compatibility conditions are exactly
Eqs.~(\ref{q1}) and (\ref{q3}). 

The conditions of applicability of the continuous approximation are discussed
in Section \ref{condition}.

\subsection{Strongly localized states}\label{strongly}

Strongly localized states with small eigenvalues correspond to the
configuration of points $x_j$ in which two points, say $x_1$ and $x_2$ are
separated by a distance $r=|x_1-x_2|$ much smaller than the mean distance 
between points $\bar{\rho} r\ll 1$.
Let $u_1$ and $u_2$ be the eigenvector components    at
points $x_1$ and $x_2$. The eigenvalue equation (\ref{4}) gives
\begin{equation}
\Lambda u_1=ru_2 +\sum_{j\neq 1,2}|x_1-x_j|\ u_j,\;\;\;
\Lambda u_2=ru_1 +\sum_{j\neq 1,2}|x_2-x_j|\ u_j.
\end{equation}
As $x_1\approx x_2$ the sums above are approximately equal and by subtracting
these equations one obtains that  to leading order in $\bar{\rho}r$
\begin{equation}
\Lambda =-r.
\label{r}
\end{equation}
Starting from this value of $\Lambda$ it is possible to
build a perturbation theory in higher powers of $\bar{\rho}r$.  

Therefore, each time that there exists two points anomalously close
to each other,
a strongly localized state with the eigenvalue (\ref{r}) is formed.
The density of such states is equal to the probability that
two uncorrelated  points are separated by a small distance $r=-\Lambda$.
From Eq.~(\ref{poisson}) it follows that at negative 
$\lambda=\bar{\rho} \Lambda $
\begin{equation}
\rho(\lambda)\sim e^{\lambda}.
\label{localized}
\end{equation}
Strictly speaking Eq.~(\ref{localized}) is applicable only for very small
$\lambda$ and it indicates that in the one-dimensional case $\rho(\lambda)$ tends
to a constant when $\lambda \rightarrow 0$.

A convenient way to distinguish between localized and delocalized states is
to compute the  participation ratio ${\bf R}$ 
\begin{equation}
{\bf R}=\frac{(\sum_{j=1}^N u_j^2)^2}{\sum_{j=1}^N u_j^4}.
\label{PR}
\end{equation}
When an eigenfunction is delocalized, all $u_j$ are of the same order and
the participation ratio  $N$ increases linearly with $N$, ${\bf R}\sim N$.
For strongly localized states the
participation ratio is independent of $N$ and ${\bf R}\sim l_{loc}$.
Therefore, when  the number $N$ of points is fixed, the participation ratio 
as  function of the corresponding eigenvalue is a  
constant (proportional to $N$) till it becomes equal to the localization
length and  ceases to depend on $N$. 

\begin{figure}
\begin{center}
\epsfig{file=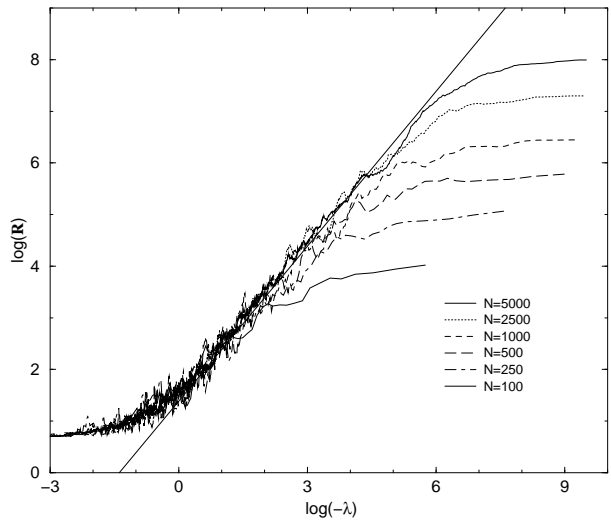, angle=-90, width=8cm}
\end{center}
\caption{Participation ratio corresponding to a single realization for the
  unit interval smoothed over a small window of $\delta \lambda$   with 
  different number $N$  of points. Straight line: asymptotics (\ref{asloc})
  of the localization length,  for comparison.}
\label{fig1}
\end{figure}

On  Fig.~\ref{fig1} the results for  the participation ratio from numerical
simulations are displayed and the expected behaviour is clearly seen.
${\bf R}$ is constant and proportional to $N$  far from the origin (on the 
right hand side of the figure), and curves corresponding to
different $N$ coalesce to the localization length  when approaching
the origin (towards the left of the figure).   

\section{Distance matrices on the circle}\label{matcircle}

The distance between two points $(i$ and $j)$ on a manifold is defined as the length of
the shortest geodesic connecting them. For later discussion, it is
convenient to distinguish between two kinds of manifolds. We shall call
a manifold for which there exists a single geodesic connecting
two points a manifold of the first kind and of the second kind otherwise. 
The simplest examples of this classification are provided by the interval
and the circle for the first and second kind respectively. 

Let us proceed to discuss the case of the circle. For a circle of radius $R$
parameterized by the polar angle $\varphi$  the distance is
\begin{equation}
||\varphi_i-\varphi_j||=R\left \{\begin{array}{ll}
|\varphi_i-\varphi_j|,&\;\mbox{if}\;|\varphi_i-\varphi_j|\leq \pi\\
2\pi-|\varphi_i-\varphi_j|,&\;\mbox{if}\;\pi<|\varphi_i-\varphi_j|\leq 2\pi
\end{array}\right . .
\label{distcircle}
\end{equation}
This equation differs from Eq.~(\ref{distline}) and the arguments of the previous
Section must be slightly modified, in particular for the crystal configuration
consisting of $N$ equally spaced points located at  $\varphi_j=2\pi j/N$.
In this case  the distance matrix $M$ takes the form
\begin{equation}
M_{ij}=||\varphi_i-\varphi_j||=\frac{2\pi R }{N}f(i-j),\;\;\;\;i,j=1,\ldots,N,
\label{circle}
\end{equation}
where $f(k)$ are the integers
\begin{equation}
f(k)=\left \{\begin{array}{ll} k,&0\leq k\leq [N/2]\\
N-k,& [N/2]<k<N \end{array}\right . 
\end{equation}
and $[x]$ is the integer part of $x$. $M$ is therefore a circulant matrix
whose successive rows are obtained by cyclic permutations of the first one 
(see e.g. \cite{Mehta}). Its eigenvectors are the Fourier harmonics
\begin{equation}
u_j^{(n)}=e^{2\pi i nj/N},\;\;\;\; n,j=1,\ldots,N
\end{equation}
with eigenvalues
\begin{equation}
\Lambda_n=\frac{2\pi}{N}\sum_{k=1}^Nf(k)e^{2\pi i nk/N}.
\label{int}
\end{equation}
As in the preceding Section it is convenient to define the
scaled eigenvalues $\lambda$ as
\begin{equation}
\lambda=\Lambda \frac{N}{2\pi R}.
\end{equation}
The sum (\ref{int}) takes a different form  for $N$ even or odd.
For $N$  even
\begin{equation}
\lambda_n=-\frac{1-(-1)^n}{2\sin^2(\pi n/N)},\;\;\;\;\;n=1\ldots,N-1,\;\;\;\;\;
\lambda_N=\frac{N^2}{4},
\end{equation}
and for $N$ odd
\begin{equation}
\lambda_n=-\frac{1-(-1)^n\cos(\pi n/N)}{2\sin^2(\pi n/N)},
\;\;\;\;\;n=1\ldots,N-1,\;\;\;\;\;
\lambda_N=\frac{N^2-1}{4}.
\end{equation}
The eigenvalues with $n$ and $N-n$ are degenerate due to the fact
that both $u_j^{(n)}$ and $u_j^{(n)*}=e^{-2\pi i nj/N}$ are eigenvectors of
the distance matrix (\ref{circle}). When $n/N\ll 1$
\begin{equation}
\lambda_n=\lambda_{N-n}\rightarrow 
-(1-(-1)^n)\frac{N^2}{2\pi^2 n^2},
\label{smallncircle}
\end{equation}
similar to Eq.~(\ref{smalln}) for the crystal solution for the interval
except for  exact two-fold degeneracies  for the circle. 

As for the distance matrix on the interval, the asymptotic behaviour of
eigenvalues  with $n/N\rightarrow 0$ for the circle can be calculated 
by considering $u_j$ as a continuous function
$u(2\pi j/N)$. In this approximation the eigenvalue equation reads
\begin{equation}
\Lambda u(\varphi)=\frac{N R}{2\pi }\int_0^{2\pi}
||\varphi-\varphi '||\;u(\varphi ')d\varphi '.
\label{52}
\end{equation}
Taking the second derivative one gets
\begin{equation}
  \Lambda u''(\varphi)=\frac{N R}{\pi}(u(\varphi)-u(\varphi-\pi)).
\end{equation}
The second term appears due to the definition (\ref{distcircle}) of
the distance on the circle when $|\varphi-\varphi '|>\pi$.
The periodic solutions of this equation are $e^{\pm  i n \varphi }$ with 
eigenvalues given by (\ref{smallncircle}). Notice that for $n$ odd 
the contribution to Eq.~(\ref{52}) from angles close to $\pi $ is the same as the 
contribution from small angles and for $n$ even they cancel each other.

\subsection{Strongly localized states}\label{stronglycircle}

The behaviour of the eigenvalue 
density for the distance matrix on the circle is practically the same as on 
the interval (except for  quasi-degenerate doublets). However, strongly
localized eigenfunctions for the circle differ from those for the interval. 
This is in contrast to the familiar situation 
(e.g. for the Anderson model) in which strongly localized
wave functions do not depend on the choice of boundary conditions. The
origin of this difference is to be found in the (unusual) growth 
of the matrix elements with the distance.

Let us assume that  an eigenfunction is localized in a region $L$ of
the order of the localization length $l_{loc}$ with $l_{loc}\ll 1$ and
$u_i$ are (large) components of this eigenfunction inside this region.
Consider a certain point $\varphi_0$ (measured from a point inside $L$) at
a distance large in comparison with the size of the localization region.
Due to localization the value $u_0$ of the eigenfunction at this point
decreases exponentially $|u_0|\sim e^{-|\varphi_0/l_{loc}|}$.
On the other hand $u_0$ has to be computed from Eq.~(\ref{eigenvalue}) 
where the sum can be restricted  to points lying in  
the localization region
\begin{equation}
\Lambda u_0= \sum_{i\in L}||\varphi_0-\varphi_i||\;u_i.
\label{eigencircle}
\end{equation}
Let $0<\varphi_0-\varphi_i<\pi$.  Then
\begin{equation}
\Lambda u_0= \varphi_0\sum_{i\in L} u_i-
\sum_{i\in L} \varphi_i \; u_i.
\label{standard}
\end{equation}
As the eigenfunction considered is a localized
state,  $|u_0|$ should be much smaller than $|u_i|$  for all
$\varphi_0\gg l_{loc}$. But the sums on the right hand side
include only the values of $u_i$ inside $L$ which are (almost) independent of
$\varphi_0$.  Therefore, in order to obey the
localization property, the  $u_i$'s inside the localization region 
should satisfy
\begin{equation}
\sum_{i\in L} u_i\approx 0,\;\;\;\sum_{i\in L} \varphi_i\ u_i \approx 0.
\end{equation}
Here the sign $\approx 0$ means that these sums
should be exponentially small ($\sim e^{-|\varphi_0|/l_{loc}}$). When
only powers of  $l_{loc}/\varphi_0$ are considered, the above sums 
are zero
\begin{equation}
\sum_{i\in L} u_i=0,\;\;\;\sum_{i\in L} \varphi_i\ u_i = 0,
\label{multipole}
\end{equation}
which can be interpreted as conditions for the vanishing of the total charge
and the total dipole moment of charges  $u_i$ located at $\varphi_i$.
These are the only general relations to be satisfied for the interval.
They do not depend on $\varphi_0$ and express the  necessary conditions for the
vanishing of the eigenfunction outside the localization region.

However, for the circle, a new feature appears when the point 
$\varphi_0$ is close to
a region diametrically opposite to the localization region $L$.
In that region  $\varphi_0=\pi+\psi_0$ with
$|\psi_0|\ll 1$ and, due to the definition of the distance
(\ref{distcircle}), Eq.~(\ref{eigencircle}) takes the form
\begin{equation}
\Lambda u_0= \pi\sum_{i\in L} u_i-
\sum_{\varphi_i <\psi_0} (\psi_0-\varphi_i) u_i + \sum_{\varphi_i>\psi_0} 
 (\psi_0-\varphi_i)   u_i.
\label{opposite}
\end{equation}
The important difference with Eq.~(\ref{standard}) is that the right
hand side of this equation depends strongly on $\varphi_0$ which determines
the splitting between negative and positive sums in (\ref{opposite}). 
No simple conditions can be imposed on the values of the eigenfunction inside
the localization region (similar to Eqs.~(\ref{multipole})) insuring
naturally that the left hand side of Eq.~(\ref{opposite}) is small. 
Consequently, our assumption (required in the usual localization theory)
that a circle eigenfunction is localized only in one small region is
not correct and the above arguments indicate that eigenfunctions of  distance
matrices on the circle are, in general,  localized not in one
but  in  at least two diametrically opposite regions. 
Exceptions to this rule may be constituted by 
states localized in such a  small region that the diametrically opposite 
one contains no points (i.e. one of the sums in Eq.~(\ref{opposite}) is
empty).

\begin{figure}
\begin{center}
\epsfig{file=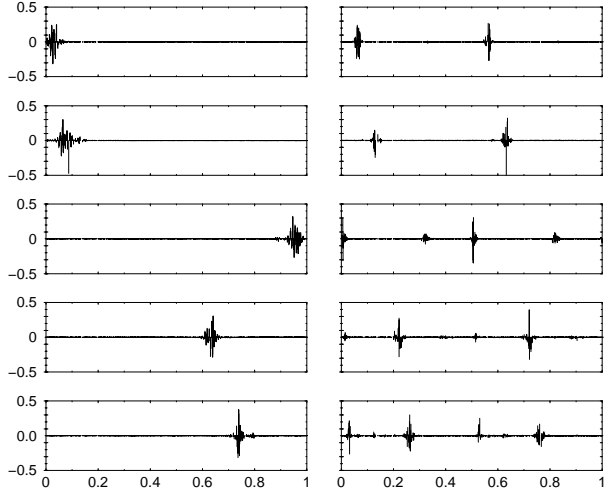, angle=-90, width=8cm}
\end{center}
\caption{Individual eigenfunctions $u^{(n)}$ corresponding to the
  $n$-th eigenvalue  with $N=1000$ points  on  the unit interval 
  (left hand side) and on the unit 
  circle (right hand side). From top to bottom: $n=$ 200,
  210, 220, 230, and 240.}
\label{fig2}
\end{figure}

On Fig~\ref{fig2} numerically calculated eigenfunctions of the distance
matrices for the  interval and the circle are plotted. 
In both figures the abscissa axis is the distance from the origin divided by
the total length. As predicted, for the
case of the interval (left hand side) each eigenfunction is localized in one small region
whereas for the circle (right hand side) the eigenfunctions are large in
two diametrically opposite regions (regions whose abscissas differ by a value
of $1/2$).

Later we will show that 
the sort of `echo' discussed here is  present, in general,  for
distance  matrices on manifolds  of the second kind. Examples are given in
the next Section.

\section{Higher-dimensional spaces}\label{higher}
In this Section we generalize the methods developed for the  one-dimensional case
to higher-dimensional spaces.

\subsection{Continuous approximation}\label{continuous}

The asymptotics of the average eigenvalue density at large negative
eigenvalues is related to delocalized states whose contribution can be
calculated in the continuous approximation. It is then necessary 
to solve  the following equation
\begin{equation}
\Lambda u(\vec{x})=\frac{N}{V}\int_{X}||\vec{x}-\vec{y}||\;u(\vec{y})d \vec{y},
\label{local}
\end{equation}
where the points $\vec{x}$ and $\vec{y}$ belong to a $d$-dimensional base manifold
$X$ of volume $V$. 

In a small vicinity of each regular point the manifold can be considered as
a part of the $d$-dimensional Euclidean space $R^d$ with coordinates
$\vec{z}$. In such vicinity  eigenfunctions of Eq.~(\ref{local}) can be
considered as functions of $\vec{z}$ and we seek for  semiclassical-type
solutions 
\begin{equation}
u(\vec{z}\;)\sim e^{i\vec{q}\vec{z}}
\label{semiclassical}
\end{equation}
with a large  vector $\vec{q}$.

Locally Eq.~(\ref{local}) leads to the following expression for the
eigenvalues
\begin{equation}
\Lambda (q)\approx \frac{N}{V}\int_{R^d} |\vec{z}|e^{i\vec{q}\vec{z}}d\vec{z}.
\label{lg}
\end{equation}
This formula is valid for manifolds of the first kind where two points
can be connected by a single geodesic. For manifolds of the second kind
(like spheres) there exist a few regions which will contribute to
$\Lambda (q)$. For clarity only the first case will be considered in detail.

Formally the integral (\ref{lg}) is divergent but it can be computed 
from the  convergent integral 
\begin{equation}
I(\alpha, \vec{q}) =\int_{R^d} e^{-\alpha
  |\vec{z}|}e^{i\vec{q}\vec{z}}d\vec{z}
\;\;\;\;\;(\alpha >0),
\label{integral}
\end{equation}
by using
$ \Lambda (q) =-(N/V) \partial I(\alpha, \vec{q})/\partial \alpha|_{\alpha=0}$.
The integral (\ref{integral}) can be expressed through the Beta function
(see e.g. \cite{Bateman} V. 1, 1.5.1) and the final result is
\begin{equation}
\Lambda (q) =-\Omega_{d-1}(d-1)!\frac{N}{V}(q^2)^{-(d+1)/2},
\label{lambdaq}
\end{equation}
where $\Omega_{d-1}=2\pi^{d/2}/\Gamma (d/2)$ is the volume of the 
$(d-1)$-dimensional unit sphere  $x_1^2+\ldots+x_d^2=1$,
and  $\Gamma (x)$ is  the Gamma function. 

When the base manifold is a part of the $d$-dimensional Euclidean space and $d$
is  odd, the result (\ref{lambdaq}) can also be obtained by successive 
differentiation of both sides of Eq.~(\ref{local}) similar to what  was done
in Section \ref{line}. In this case one also obtains  exact relations
between  the eigenfunctions of the distance matrix and those of the
Laplacian for $d=1$, bi-Laplacian for $d=3$, etc. 

For any smooth boundary conditions  the density  of solutions of
the form (\ref{semiclassical}) is  asymptotically the same as for the
spectrum of the Laplacian $(\Delta +q^2)\Psi=0$, given by
\begin{equation}
\rho(q)=V\int_{R^d} \frac{d \vec{k}}{(2\pi)^d}
\delta(q-|\vec{k}|)=\frac{V\Omega_{d-1}}{(2\pi)^d}q^{d-1},
\label{densityn}
\end{equation}
where $V$ is the volume of the manifold. 

From Eqs.~(\ref{lambdaq}) and (\ref{densityn}) the following  estimate 
of the tail
of the integrated density of eigenvalues of distance matrices is obtained
\begin{eqnarray}
{\bf N}(\Lambda)&\approx&
\frac{1}{N}\int_0^{\infty} \Theta(\Lambda -\Lambda (q))\rho (q)dq\nonumber\\
&=&\frac{V\Omega_{n-1}(q(\Lambda))^{d}}{(2\pi)^d N\ d}=
C_d\left (\frac{N}{V}\right )^{1/(d+1)}(-\Lambda)^{-d/(d+1)},
\end{eqnarray}
where $q(\Lambda)$ is the inverse of the function $\Lambda (q)$ defined  in
Eq.~(\ref{lambdaq})  
and $C_d$ is a constant depending only on the dimensionality of the system.

Introducing the dimensionless scaled  eigenvalues
\begin{equation}
\lambda=\Lambda (\frac{N}{V})^{1/d},
\end{equation}
this result can be rewritten in the universal form
\begin{equation}
{\bf N}(\lambda) \approx C_d(-\lambda)^{-d/(d+1)},
\label{nl}
\end{equation}
where ${\bf N}(\lambda)$ is the counting function in the variable $\lambda$.

For manifolds of the second kind like spheres the only
modification of the above results is a slight change of the scaled eigenvalue 
\begin{equation}
\lambda=\Lambda (\frac{gN}{V})^{1/d},
\end{equation}
where $g$ is the number of singular regions contributing to Eq.~(\ref{lg}).
For spheres of arbitrary dimensions $g=2$. 

\begin{figure}
\begin{center}
\epsfig{file=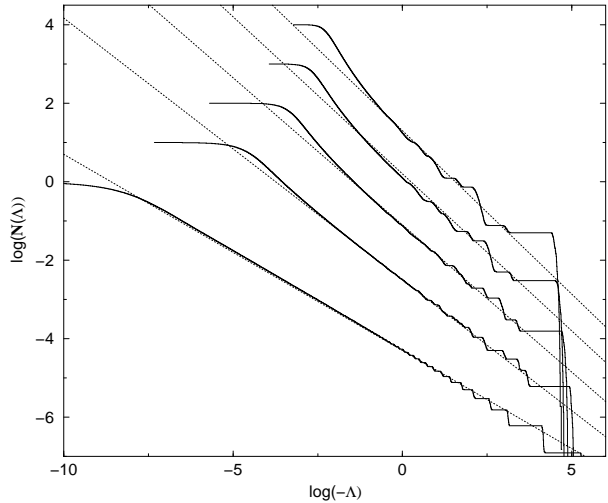, angle=-90, width=8cm}
\end{center}
\caption{Averaged (over 50 realizations) staircase function with $N=1000$
  points  in  the unit hyper-cube of dimension $d=1,2,3,4,5\;$ from
  bottom to top.
  Straight dotted lines of slope $-d/(d+1)$, as predicted by Eq.~(\ref{nl}),
  for comparison. For clarity curves are shifted vertically by $d-1$  units.}
\label{fig3}
\end{figure}

On Fig.~\ref{fig3} results of numerical calculations of the
average staircase function for hyper-cubes of different dimensions are
compared with  the prediction (\ref{nl}). They are in very good agreement.

\subsection{Quasi-multiplets}\label{multiplets}

The estimates of the previous Section are general but they do not take
into account the fine structure of the eigenfunctions.

Let us assume that the manifold $X$ is invariant
under a certain symmetry group $G$. As the kernel $||\vec{x}-\vec{y}||$ 
of Eq.~(\ref{local}) is the distance between two points on this manifold,
it remains unchanged under simultaneous transformation of $\vec{x}$ and
$\vec{y}$. Therefore from Eq.~(\ref{local}) it follows that the transformed
eigenfunction
\begin{equation}
u'(\vec{x})=u(G(\vec{x}))
\end{equation}
is also a solution of this equation. From this simple remark it is clear
that in the
continuous approximation the eigenfunctions $u(\vec{x})$ form  irreducible
representations of the symmetry group of the initial manifold similar to
solutions of the Laplace equation. When this
group has $h$-dimensional irreducible representations the eigenvalues of
distance matrices will be $h$-times degenerate.  

To be specific let us consider in detail the case of the $d$-dimensional sphere 
as the base manifold. The invariance group of the
sphere is the $d$-dimensional rotation group.  Let $p=d-1$. It is well known 
(see e.g. \cite{Bateman} V. 2, 11.2) that 
the harmonic polynomials   of degree $p+2$ (hyper-spherical harmonics)
$Y_{l  \vec{m}}(\vec{\theta}, \varphi)$ form the basis of irreducible
representations of the rotation group. Here 
$\vec{\theta}=\theta_1,\ldots, \theta_p$ and  $\varphi$
are the standard hyper-spherical angles and $\vec{m}=m_1,\ldots, m_p$ are
integers obeying the  inequalities
\begin{equation}
0\leq m_p \leq \ldots \leq m_1\leq p+2.
\end{equation}
The dimensions $h(l,p)$ of these  representations are
\begin{equation}
h(l,p)=(2l+p)\frac{(l+p-1)!}{p!l!}
\end{equation}
(for $d=2$, $h(l,1)=2l+1$, and $Y_{l m}(\theta, \varphi)$ are the usual spherical
harmonics).

The  eigenvalues corresponding to these eigenfunctions have multiplicity
$h(l,p)$. Their explicit form can  easily be derived directly from
the invariance of Eq.~(\ref{local}) under rotation. Choose the $z$-axis along
the vector $\vec{x}$. Introducing the hyper-spherical coordinates in the
usual way one concludes that $u(\vec{y})$ equals the unique harmonic
polynomial
depending only on $\cos\theta$ where $\theta$ is the angle between vectors
$\vec{x}$ and $\vec{y}$ which is proportional to the Gegenbauer polynomial 
$C_l^{p/2}(\cos\theta)$ (see e.g. \cite{Bateman} V. 2, 11.2).
Therefore
\begin{equation}
\Lambda_l=C_{lp}\frac{N}{V}\int_0^{\pi}\theta C_{l}^{p/2}(\cos \theta)\sin^p\theta d
\theta,
\end{equation}
with a constant $C_{lp}$ depending on $p$ and $l$. The explicit form of 
$\Lambda_l$ is not instructive for our purposes.

From properties of the Gegenbauer polynomials (see e.g. \cite{Bateman} V.
2, 10.9) it follows that all $\Lambda_l$ with even $l\neq 0$ are zero and
consequently  beyond the reach of  the continuous approximation (see Section
\ref{condition}). The value
corresponding to $l=0$ is the Perron-Frobenius eigenvalue. We therefore
concentrate on odd values of $l$.

\begin{figure}
\begin{center}
\epsfig{file=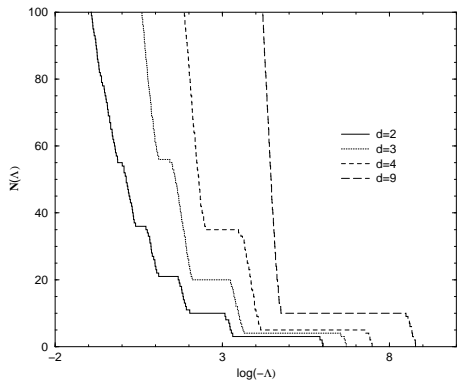, angle=-90, width=6cm}
\end{center}
\caption{Averaged staircase function with $N=1000$ points 
  on the $d$-dimensional  unit sphere ($d=2,3,4,9$) showing quasi-degenerate multiplets.
  For clarity, the counting function has been  normalized to $N$.}
\label{fig4}
\end{figure}

In the continuous approximation the eigenvalues
for $d$-dimensional sphe\-res are $h(2k+1,d-1)$ times degenerate. For the
one-dimensional sphere
(i.e. the circle) $h(2k+1,0)=2$ as has been seen in Section \ref{matcircle}.
For the 2-sphere
(the usual sphere) $h(2k+1,1)$ equals $3,7,11\ldots$, for the 3-sphere the
first multiplicities are $4, 16, 36,\ldots$, for the 4-sphere they are
$5,30,91,\ldots,$ etc. In general the first multiplet for the $d$-sphere
corresponding to $l=1$ has multiplicity $d+1$, i.e. it is equal to the dimension
of the embedded space.
On Fig.~\ref{fig4} these degeneracies can be read off from the numerically calculated
average staircase function for spheres of different dimensions.

\begin{figure}
\begin{center}
\epsfig{file=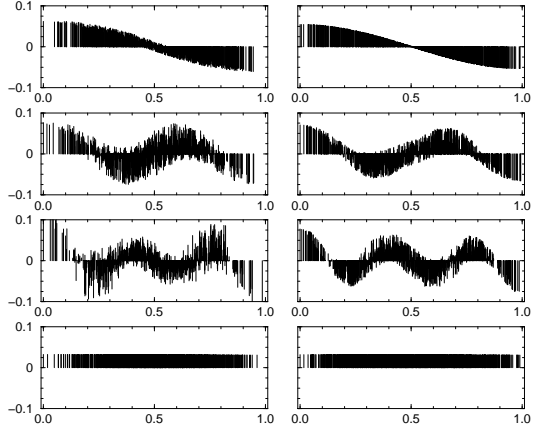, angle=-90, width=7cm}
\end{center}
\caption{Individual eigenfunctions $u^{(n)}$ corresponding to the
  $n$-th eigenvalue with $N=1000$
  points on the unit 3-sphere (left hand side) and on the unit
  2-sphere (right  hand side). From top to bottom: $n=1,5,$ and $21$ for
  $d=3$  and $n=1,4,$ and
  $11$ for $d=2$. These values of $n$  correspond to the lowest eigenvalue  of each of the first
  three quasi-multiplets. The Perron-Frobenius eigenfunction ($n=1000$) is at
  the bottom. See text for further explanation.}
\label{fig4b}
\end{figure}

We also show on Fig.~\ref{fig4b} the structure of some individual
eigenstates. We proceed as follows. Given an eigenvector $u$, choose its
largest (in absolute value) component, say $u_1$, and plot the components
$u_j$ as a function of the distance  $||\vec{x}_1-\vec{x}_j||$ (normalizing
to the maximum distance between two points on the manifold).  
By construction, the highest component corresponds to the zero
value of the abscissa. For the 1-dimensional case, this procedure
corresponds to the natural ordering of points on the interval or the circle.
The quasi-multiplet structure of the eigenstates is clearly visible on the
figure. For
comparison, the Perron-Frobenius eigenstate is also displayed. For spheres its
components are constant to within $1/\sqrt{N}$ fluctuations.

\subsection{Condition for  applicability of the continuous
  approximation}\label{condition}

The continuous approximation is based on the well known fact that under
quite general conditions the sum of a large number of independent random
variables $\vec{x}_j$ with distribution $d\mu(\vec{x})$ tends to its mean
value
\begin{equation}
\frac{1}{N}\sum_{j=1}^Nf(\vec{x}_j)=\int
f(\vec{x})d\mu(\vec{x})+\frac{\zeta}{\sqrt{N}},
\label{ergodic}
\end{equation}
where, when $N\to \infty$, $\zeta$ is a random variable with zero mean and 
variance independent on $N$. 

This type of `ergodic theorem' makes  natural to consider, instead of
the true eigenvalue Eq.~(\ref{eigenvalue}), 
\begin{equation}
\tilde{\Lambda}_n u_{i}^{(n)}=\frac{1}{N-1}
\sum_{j=1}^{N}||\vec{x}_i-\vec{x}_j|| \; u_j^{(n)},
\label{true}
\end{equation}
its continuous approximation
\begin{equation}
\tilde{\Lambda}_n^{(c)}u_n(\vec{x}\ )=
\frac{1}{V}\int_X ||\vec{x}-\vec{y}||\; u_n(\vec{y}\ )d\vec{y}
\label{ca}
\end{equation}
(we introduce for convenience the factor $1/(N-1)$ with the
corresponding redefinition $\tilde{\Lambda}_n=\Lambda_n/(N-1)$ because
in the sum in (\ref{true}) there are only $N-1$ non-zero terms).
Here for simplicity uncorrelated points $\vec{x}_j$ uniformly 
distributed on the  base manifold $X$ (i.e. $d\mu(\vec{x})=d\vec{x}/V$) are
considered.

As the kernel of the integral equation (\ref{ca}) is symmetric, its
eigenvalues $\tilde{\Lambda}_n^{(c)}$ are real and its eigenfunctions
$u_{n}(\vec{x}\ )$ can be chosen real orthogonal
\begin{equation}
\frac{1}{V}\int_X u_{n}(\vec{x}\ ) u_{m}(\vec{x}\ )d\vec{x}=\delta_{mn}.
\label{norm}
\end{equation}
Let us first consider the case when the eigenvalue
$\tilde{\Lambda}_n^{(c)}$ of the continuous equation (\ref{ca})  is non
degenerate and let us look for solutions of the true eigenvalue equation
(\ref{true}) in the form 
\begin{equation}
u_j^{(n)}=u_n(\vec{x}_j)+\sum_{m\neq n} c_m u_m(\vec{x}_j),
\end{equation}
where $c_m$ are considered  as small quantities. (This procedure is often used
in perturbation theory when dealing with the Schroedinger equation.)
Substituting these expressions in Eq.~(\ref{true}), multiplying  both sides  
by $u_n(\vec{x}_i)$, and summing from $i=1$ to  $N$ one obtains
\begin{eqnarray}
&&\tilde{\Lambda}_n\frac{1}{N}\sum_{i=1}^Nu_n(\vec{x}_i) 
\left (u_n(\vec{x}_i)+\sum_{m\neq n} c_m u_m(\vec{x}_i)\right )=
\nonumber\\
&&\frac{1}{N(N-1)}\sum_{i,j=1}^Nu_n(\vec{x}_i)\;||\vec{x}_i-\vec{x}_j||
\left (u_n(\vec{x}_j)+\sum_{m\neq n} c_m u_m(\vec{x}_j)\right ).
\label{inter}
\end{eqnarray}
The mean values of the terms which are multiplied by $c_m$ are zero and,
consequently, they are of the order of $1/\sqrt{N}$ (see Eq.~(\ref{ergodic})).
Therefore in Eq.~(\ref{inter}) these terms can be ignored 
to  first order in $1/\sqrt{N}$ and this equation is reduced  to 
\begin{equation}
\tilde{\Lambda}_nA_N(n)=B_N(n),
\end{equation}
where
\begin{eqnarray}
A_N(n)&=&\frac{1}{N}\sum_{i=1}^Nu_n(\vec{x}_i) u_n(\vec{x}_i),\\
B_N(n)&=&\frac{1}{N(N-1)}\sum_{i,j=1}^Nu_n(\vec{x}_i)\;||\vec{x}_i-\vec{x}_j||\;
u_n(\vec{x}_j).  
\end{eqnarray}
When $N\to \infty$ these sums  tend to their mean value plus corrections
\begin{equation}
A_N(n)\to 1+\frac{\sigma (n)}{\sqrt{N}},\;\;\;
B_N(n)\to 
\tilde{\Lambda}_n^{(c)}+\frac{\Sigma (n)}{\sqrt{N}}.
\end{equation}
The mean value of $\sigma (n)$ and $\Sigma (n)$ is zero.
When $N\to \infty$ their variances are independent on $N$ and
can be computed from straightforward calculations.

By taking the average of  $A_N(n)$ and $B_N(n)$ one recovers the continuous
approximation result. The first correction 
$\delta \Lambda_n=\tilde{\Lambda}_n-\tilde{\Lambda}_n^{(c)}$
is determined by their fluctuating part
\begin{equation}
\delta \Lambda_n=\frac{1}{\sqrt{N}}
(\Sigma(n)-\tilde{\Lambda}_n^{(c)}\sigma (n)).
\end{equation}
The mean value of $\delta \Lambda_n$ is, of course, zero and 
its variance is equal to
\begin{equation}
<\delta \Lambda_n^2>=\frac{\tilde{\Lambda}_n^{(c)\ 2}}{N}v^2(n),
\label{splitting}
\end{equation}
where 
\begin{equation}
v^2(n)=\int_X u_n^4(\vec{x}\ )d\mu(\vec{x})-1,
\end{equation}
provided that the $u_n(\vec{x}\ )$ are orthogonal.
For semiclassical-type solutions (\ref{semiclassical}), where real orthogonal solutions can be
chosen in the form $\sqrt{2}\sin \vec{q}\vec{z}$ and
$\sqrt{2}\cos \vec{q}\vec{z}$, $v^2(n)=5$, independent of $n$.

If the eigenvalue $\tilde{\Lambda}_n^{(c)}$ of the continuous equation (\ref{ca}) 
is $h$-fold degenerate,  a simple extension of the 
previous arguments leads to an estimate of the splitting of the degeneracy
by including the next order corrections. 

Let $u_n^{(m)}(\vec{x}\ )$ $m=1,\ldots ,h$ be the solutions of the the continuous
equation (\ref{ca}) corresponding to an $h$-fold degenerate eigenvalue
$\tilde{\Lambda}_n^{(c)}$ normalized as follows
\begin{equation}
\int_X u_n^{(m)}(\vec{x}\ )u_n^{(m')}(\vec{x}\ )d\mu(\vec{x})=\delta_{mm'}.
\end{equation}
The total splitting 
$\Delta \Lambda_n=\sum_{j=1}^h\delta \Lambda_n^{(j)}$ is a random variable 
which has the following estimate
\begin{equation}
\frac{\Delta \Lambda_n}{\tilde{\Lambda}_n^{(c)}}=
  h\cdot \zeta \cdot \frac{v(n)}{\sqrt{N}} 
\label{corrections}  
\end{equation}
with $<\zeta>=0$ and $<\zeta^2>=1$ and
\begin{equation}
v^2(n)=\frac{1}{h^2}\sum_{m=1}^h\sum_{m'=1}^h
\int_X \left [u_n^{(m)}(\vec{x}\ )u_n^{(m')}(\vec{x}\ )\right ]^2d\mu(\vec{x})-1.
\end{equation}
More generally, if a function 
$f(\Lambda_n)$ is computed in
the continuous approximation, it will have a random (depending on
realizations) fluctuation $\delta f(\Lambda_n)$  
which to first order in $1/\sqrt{N}$ is 
\begin{equation}
\delta f(\Lambda_n) =\frac{d f(\Lambda_n)}{d\Lambda_n}h \Lambda_n \zeta
\frac{v(n)}{\sqrt{N}}. 
\end{equation}
We are  interested mainly in the counting function. According to Eq.~(\ref{nl})
it has a power-law asymptotics when smoothed over a small window 
$\delta \Lambda$ and
$d{\bf  N}(\Lambda)/\Lambda= \nu {\bf  N}(\Lambda)/\Lambda$ with
$\nu=-(d+1)/d$. Consequently
\begin{equation}
\delta {\bf N}(\Lambda) \sim {\bf N}(\Lambda)\frac{\zeta}{\sqrt{N}}.
\label{deltaN}
\end{equation}
The condition that perturbed eigenvalues do not deviate from the ones
computed in the continuous approximation (or equivalently that 
the splitting (\ref{splitting}) is  much smaller than the distance between
multiplets in the continuous approximation) leads to the following inequality
\begin{equation}
\delta {\bf N}(\Lambda)\ll \frac{1}{N}.
\label{con}
\end{equation}
Together with (\ref{deltaN}) it implies
that, in general, the continuous approximation gives correctly the
eigenvalues  of only the first $\sqrt{N}$  eigenstates ordered by
increasing eigenvalues and that by increasing $N$
more and more distinct and isolated  multiplets are present.
In terms of scaled eigenvalues, one has that  the continuous approximation
can be used for the approximation of non-averaged eigenvalues satisfying 
\begin{equation}
|\lambda|\geq  N^{(d+1)/2d}.
\end{equation}
For smaller $|\lambda|$ the spacing between successive eigenvalues computed in
the continuous approximation becomes comparable  to the fluctuating corrections
and the mixing of different states is important. 

Nevertheless, one can still use the continuous approximation for the
calculation of   average quantities (e.g. the counting function).
The main point is that
the first correction to (\ref{deltaN}) vanishes because the mean value of
$\zeta$ is zero to first order in $1/\sqrt{N}$.
If there are no singularities (which is probably true for $d>2$), 
the next order correction will be of the order
of $G(\lambda)/N$ with a certain function $G(\lambda)$ and the conditions 
(\ref{con}) on the average will be
valid till $\lambda$ is of the order of $1$. Therefore  for averaged quantities 
one can use the continuous approximation for a finite fraction  of the 
total number of eigenvalues. 

\subsection{Strongly localized states}\label{stronglygeneral}

The properties  of strongly localized states can
be estimated by slightly modifying  the arguments  used for the one-dimensional
case. 

For manifolds of the second kind (like for the circle), assume that  an
eigenfunction is localized in a small
region $L$ and $u_i$ are (large) values of this eigenfunction inside this
region. Choose the origin somewhere inside this region and consider a point
$\vec{x}_0$ at large distance from it. The value $u_0$ of the
eigenfunction  at this point from Eq.~(\ref{eigenvalue}) is
\begin{equation}
\Lambda u_0= \sum_{i \in L}||\vec{x}_0-\vec{x}_i||\; u_i.
\end{equation}
As $||\vec{x}_0||\gg ||\vec{x}_i||$, the
right-hand side of this expression can be expanded into powers of $\vec{x}_i$
\begin{equation}
\Lambda u_0= R\sum_{i \in L} u_i+
\sum_{i\in L} \vec{\zeta}\cdot \vec{x}_{i}\ u_i +\ldots,
\end{equation}
where $R=||x_0||$ and
$\vec{\zeta}=\partial ||\vec{x}_0||/\partial \vec{x}_0$.

As  $|u_0|$ should decrease exponentially with $R$ and each term in the 
above equation decreases as a different power of $R$, one concludes, similar
to the case of the circle, that   in order to obey the
localization property $u_i$ should obey the relations
\begin{equation}
\sum_{i \in L} u_i\approx 0,\;\;\;\;
\sum_{i \in L}\vec{\zeta}\cdot \vec{x}_i\ u_i \approx 0,\;\mbox{etc}.
\label{sums}
\end{equation}
In regions not too close to a region diametrically opposite to the
localization region $\vec{\zeta}=\vec{x}_0/R$ and these equations take the
form of zero multipole moments
\begin{equation}
\sum_{i \in L} u_i=0,\;\;\;\;\sum_{i\in L} \vec{x}_i u_i,\ldots
\label{multipoles}
\end{equation}
But close to the diametrically opposite region there are points for which
the distance changes its form as in Eq.~(\ref{distcircle}). For these points
the derivative of $||\vec{x}||$ will change sign and, in general, conditions
(\ref{multipoles}) will be impossible to fulfill. Consequently, the
eigenfunction in the diametrically opposite region cannot, in general, be
exponentially small as required by the usual localization theory.

\begin{figure}
\begin{center}
\epsfig{file=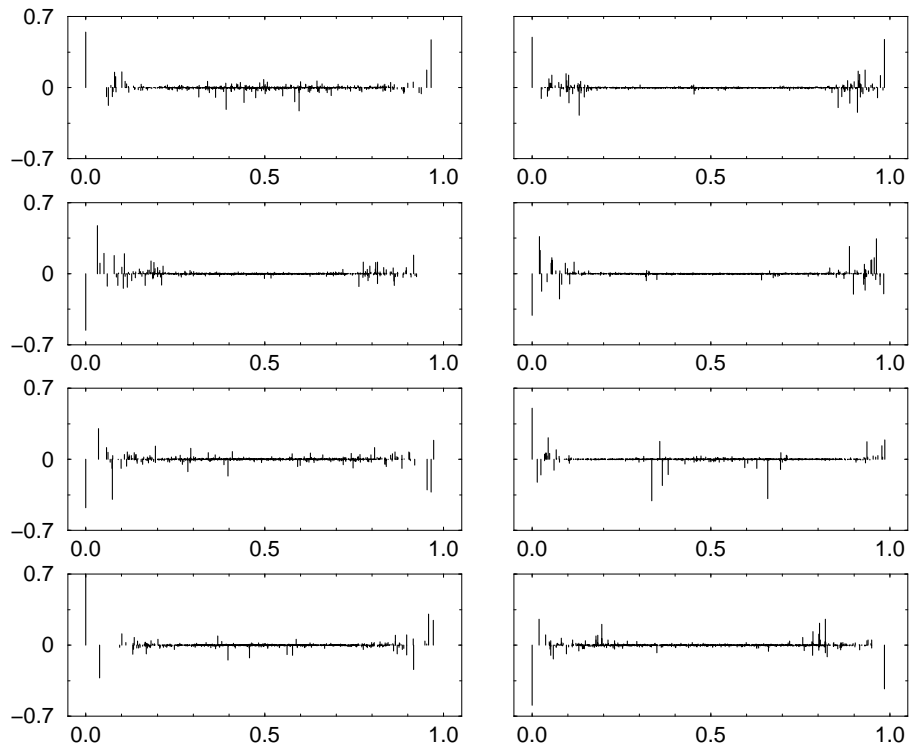, angle=-90, width=12cm}
\end{center}
\caption{Individual eigenfunctions $u^{(n)}$ as for Fig.~\ref{fig4b}
  except
  $n=$ 800, 801, 802, 803 for $d=3$ and $n=$ 650, 651, 652, 653 for $d=2$ from top
  to bottom. All states are localized. }
\label{fig8}
\end{figure}

The above  discussion  about the localization `echo' is  
valid for manifolds of the second kind (i.e. if there exist a few geodesics 
connecting two points) e.g.  for spheres of 
different dimensions. In Fig.~\ref{fig8} we present the structure of the 
localized eigenfunctions of the distance matrices for 2 and 3 dimensional
spheres in the same way as it was done in Fig.~\ref{fig4b}. 
In all cases the eigenfunctions have large values near the left and 
right ends of the interval, in  agreement with 
localization on two diametrically opposite regions. Comparison with
Fig.~\ref{fig4b} illustrates also the completely different structure of extended and
localized states (notice the difference in scale in both figures).

To compute the density of strongly localized states let us, as above,
assume that  an eigenfunction is localized in a small region $L$ and denote by
$u_i$ its large components inside $L$. (For simplicity we consider only 
manifolds of the first kind.)

The eigenvalue equation (\ref{eigenvalue}) gives
\begin{equation}
\Lambda u_i=\sum_{j\in L} ||\vec{x}_i-\vec{x}_j||\;u_j.
\end{equation}
Because we assume that all $u_i$ are localized (are large) in a small region,
all differences $||\vec{x}_i-\vec{x}_j||$ are of the order of the distance
$r$ between nearest points
\begin{equation}
\Lambda u_i\approx r \sum_{j\neq i} u_j.
\end{equation}  
As already  indicated in Eq.~(\ref{sums}), one of the necessary conditions of
localization  is
\begin{equation}
u_i+\sum_{j\neq i} u_j=0.
\end{equation}
Therefore
\begin{equation}
\Lambda \approx - r,
\end{equation}
which means that the distribution of eigenvalues of localized states is
approximately the same as the distribution of distances between two
uncorrelated points
randomly distributed on the manifold. 
For small distances the latter is proportional to the volume and,
consequently, the staircase function at small negative $\lambda$ is
\begin{equation}
{\bf N}(\lambda)\approx  1-K_d (-\lambda)^d,
\label{smallLambda}
\end{equation}
where $K_d$ is a constant depending on the dimensionality of the system.
(In situations where points though uniformly distributed exhibit
repulsion we expect that the exponent in Eq.~(\ref{smallLambda}) will be
larger.)

\begin{figure}
\begin{center}
\epsfig{file=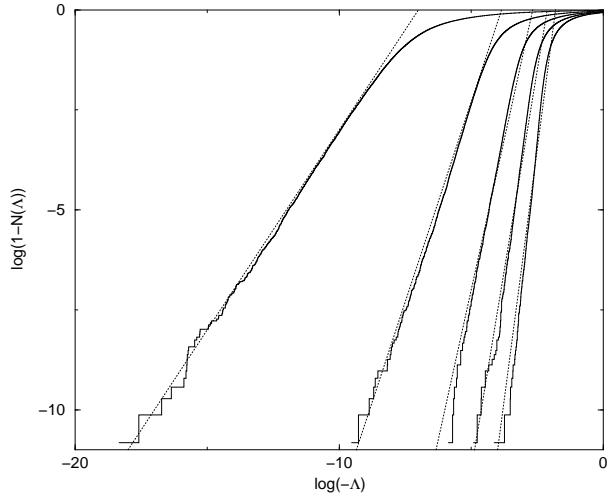, angle=-90, width=8cm}
\end{center}
\caption{Averaged fraction of eigenvalues above 
  $\Lambda$ with $N=1000$ points 
  in the unit hyper-cube of dimension 
  $d=1,2,3,4,5$ (from left to right). 
  Straight lines of  slope $d$ as predicted by
  Eq.~(\ref{smallLambda}), for comparison. }
\label{fig5}
\end{figure}

On Fig.~\ref{fig5} (same numerical data as for Fig.~\ref{fig3}) 
the small $\Lambda$ behaviour of the staircase function is displayed. It is
clear that the estimate (\ref{smallLambda}) is in  good agreement with 
the numerical simulations.

\begin{figure}
\begin{center}
\epsfig{file=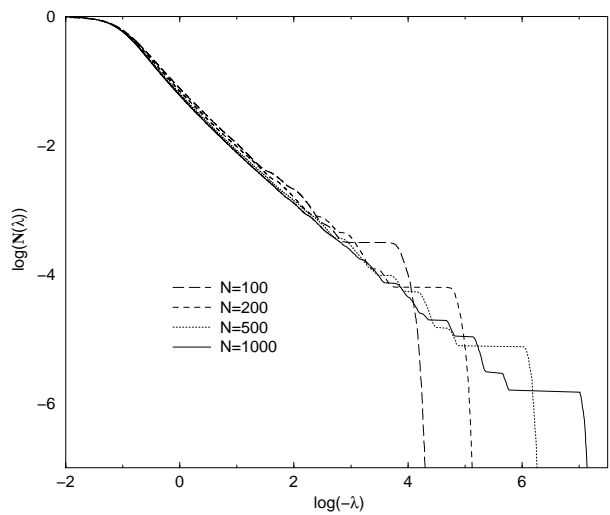, angle=-90, width=8cm}
\end{center}
\caption{ Averaged staircase function for the unit 3-dimensional cube for
  different number of points.  }
\label{fig3b}
\end{figure}

Eq.(\ref{smallLambda}) together with Eq.~(\ref{nl}), which govern the small
and large $\lambda$ behaviour of the average staircase function respectively,
strongly suggest that ${\bf N}(\lambda)$ is independent on the number of
points $N$ when $N \to \infty$. This is illustrated on Fig.~\ref{fig3b} .

\begin{figure}
\begin{center}
\epsfig{file=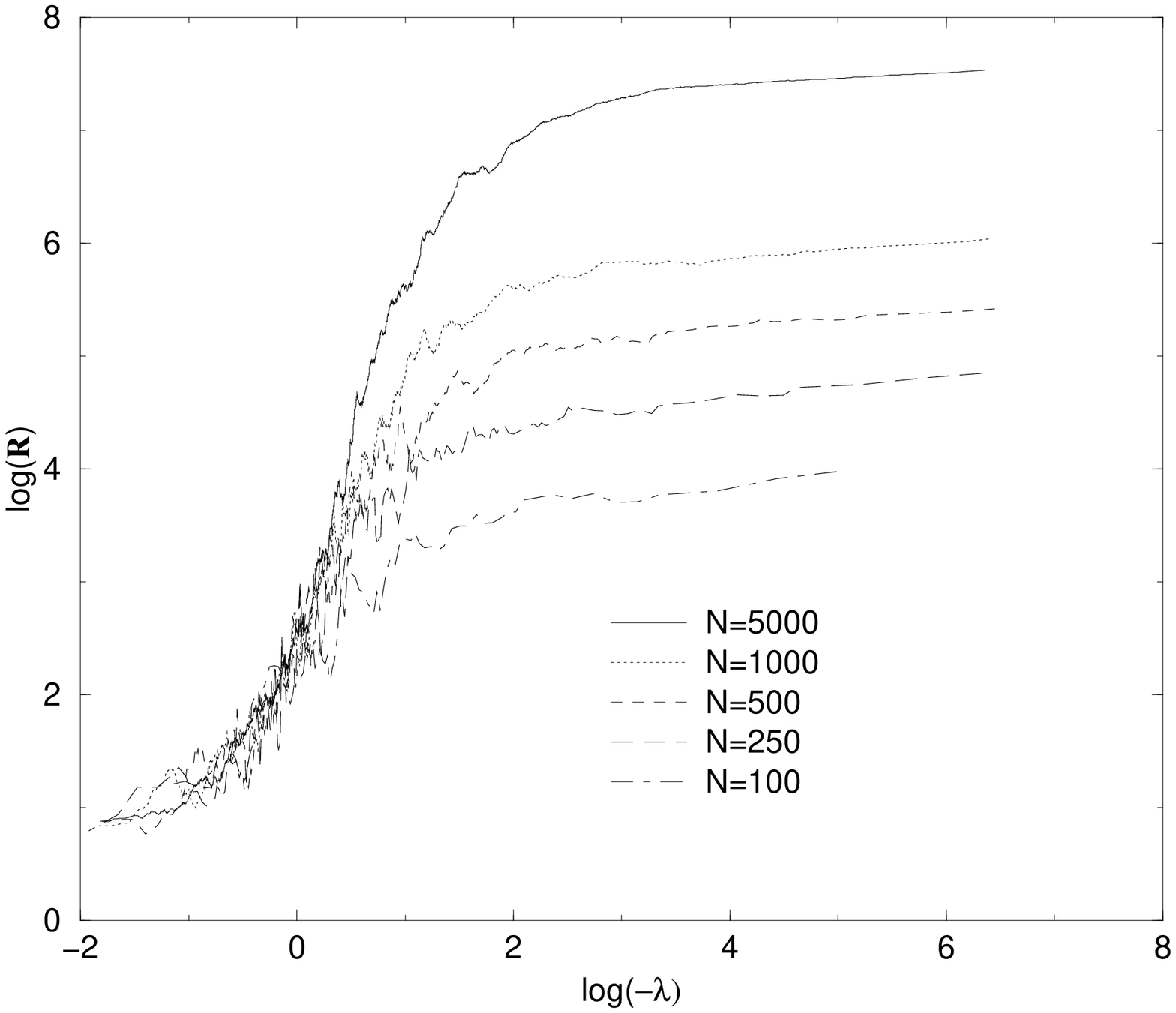, angle=-90, width=6.5cm}
\epsfig{file=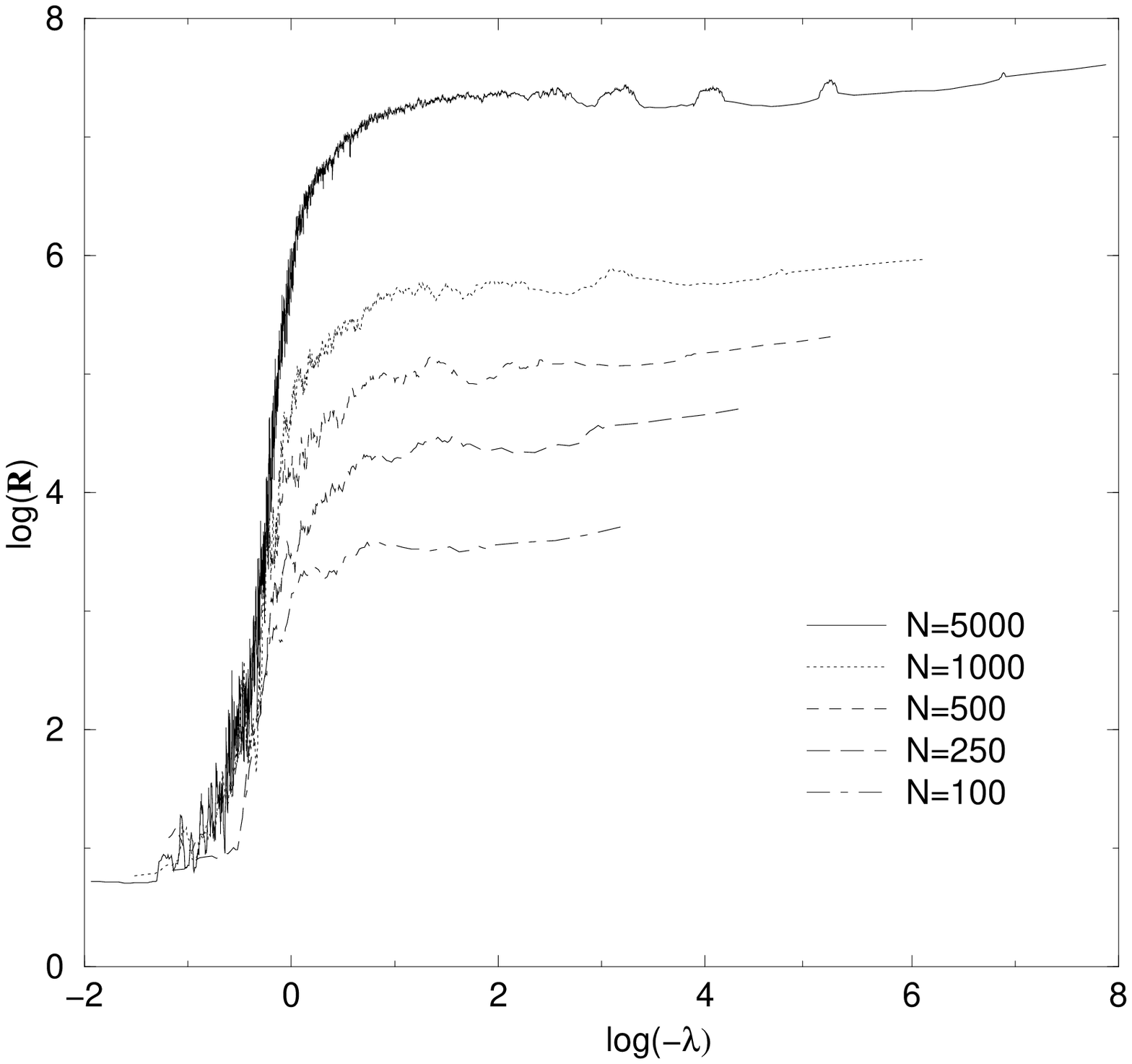, angle=-90, width=6.5cm}
\end{center}
\caption{Participation ratio corresponding to a single realization for 2-dimensional
  (left part) and  3-dimensional (right part) unit sphere smoothed over a
  small window $\delta \lambda$ with  different number $N$ of points .
  The abscissa axis is $\log(-\lambda)$ where   $\lambda=\Lambda N^{1/2}$ (left part) and
  $\lambda=\Lambda N^{1/3}$ (right part).}
\label{fig6}
\end{figure}

To get further insight on localization properties let us first remind some
qualitative properties of the Anderson model. In one dimension all states
are localized, irrespective of the size of disorder. In two dimensions
perturbation theory cannot be applied and the localization length
increases exponentially when disorder decreases. For dimensions larger than
two there is a localization-delocalization transition at finite strength of
disorder and the threshold value increases with dimensionality of the system.
For distance matrices, the parameter governing the size of the
disorder is the scaled eigenvalue $\lambda=\Lambda(N/V)^{1/d}$, with
increasing disorder as $|\lambda|\to 0$.
In Fig.~\ref{fig6} we present the numerically computed
participation ratio for spheres of dimensions 2 and 3. For the 3-sphere the
effect of the localization-delocalization transition (i.e. the sharp 
increase of the localization length at a finite value of $\lambda$) is
clearly visible. 

\section{Summary and conclusions}\label{conclusion}

Let $N$ points be  distributed on a certain manifold. The $(i,j)$
matrix element of the distance matrix is defined as  the distance between points
$i$ and $j$. Schoenberg \cite{Schoenberg} proved long time ago that 
for Euclidean manifolds all eigenvalues of such matrices, except one, are
non-positive. This property can be extended to other manifolds, in particular, 
for spherical manifolds treated here (see \cite{BBS} for a general 
discussion). 

The study of the one-dimensional  case (interval and circle) provides already 
some clues of what are the ingredients governing the properties in higher 
dimensions and the properties related to the very simple crystal configuration
are partially kept when `disorder' is added.

In general it is  demonstrated that, if the points are uncorrelated and
uniformly distributed  on a base
manifold, the average density of eigenvalues of distance matrices in the
limit $N\rightarrow \infty$ has a power-law behaviour for large and small negative
eigenvalues, the exponent  depending only on the dimension of the
manifold (see Eqs.~(\ref{nl}) and (\ref{smallLambda})). 

The eigenfunctions
of the distance matrices with large negative eigenvalues are delocalized 
(for 1 and 2 dimensional manifolds the localization length is much larger 
than the system size) whereas  the eigenfunctions with very small negative 
eigenvalues are strongly localized. 

If the manifold possess a symmetry group, large negative
eigenvalues form almost degenerate multiplets whose dimensions equal the
dimensions of the irreducible representations of the group and the
conditions for the presence of isolated multiplets are established.

A distinction among base manifolds for which  two points 
are connected by one or several geodesics is made. In the latter case 
the eigenfunctions of the distance matrix are, in general,
localized not in one but in several  regions (echo). 
For spheres of any dimension we find strongly 
localized states in two diametrically opposite regions. The understanding of
the structure of the echo for more general spaces deserves further study.

Strongly localized states are, by definition,  mostly sensitive 
to local properties of the base manifold. The existence of the echo
shows, however,  that certain global
properties of the manifold are reflected on these states as well.

What precedes is illustrated in detail by studying distance matrices of
uncorrelated points uniformly distributed on hyper-spheres and hyper-cubes 
of different dimensions. 

The introduction of distance matrices in Ref.~\cite{Vershik} was to a large
extent motivated by the fact that  they encode the metric properties of
the base manifold. 
Our results show one  possible way of solving the inverse problem, namely, 
the reconstruction of the initial manifold from the knowledge of the
spectral properties of its distance matrix. We demonstrate
that large negative eigenvalues can be approximated with $1/\sqrt{N}$ 
accuracy  by the solutions of the continuous approximation Eq.~(\ref{local}).
As this integral equation is similar  to the Laplace
equation, one may conjecture that for this equation, similarly as for the
former, the question raised by Kac ``Can one hear the shape of a
drum?''\cite{Kac}  can also be answered  affirmatively, except probably  
for very special isospectral  cases.

When  completing  this work we became aware of Ref.~\cite{Mezard} where
Euclidean Random Matrices were introduced. Matrix elements are then a
function of finite range of the relative distance between two points in
Euclidean space $M_{ij}=f(||\vec{x}_i-\vec{x}_j||)$.
One of the main differences with respect to the present work is that 
in \cite{Mezard} the results strongly depend  on the choice of $f$
whereas here it is kept fixed  $f(x)=x$ and 
emphasis is put on the choice of the manifold.  

\vspace{.5cm}

{\bf Acknowledgments} 

\vspace{.5cm}

The authors are greatly indebted to A.M. Vershik for discussing his work
\cite{Vershik} prior the publication and to M. M\'ezard for pointing out 
reference \cite{Mezard}. One of the authors (O.B.)
acknowledges hospitality at the MFI  Oberwolfach, where this
work originated.

\end{document}